\documentclass[11pt]{article}
\usepackage{moriond,epsfig}
\usepackage{pipipresen}
\def\Journal#1#2#3#4{{#1} {\bf #2}, #3 (#4)}

\def\NIMA{{\em Nucl. Instrum. Methods} A}
\def\NPB{{\em Nucl. Phys.} B}
\def\PLB{{\em Phys. Lett.}  B}
\def\PRL{\em Phys. Rev. Lett.}
\def\PRD{{\em Phys. Rev.} D}

\def\PTP{\em Prog. Theor. Phys.}

\def\be{\begin{equation}}
\def\ee{\end{equation}}
\def\bea{\begin{eqnarray}}
\def\eea{\end{eqnarray}}

\begin{document}
\vspace*{4cm}
\title{$CP$ VIOLATION IN $\Btopipi$ DECAY}

\author{T. NAKADAIRA\\(for Belle collaboration)}

\address{Department of Physics, Faculty of Science, the University of Tokyo,\\
7-3-1 Hongo, Bunkyo-ku, Tokyo 113-0033, Japan}

\maketitle\abstracts{%
  We present an improved measurement of $CP$-violating asymmetries in
  $B^0 \rightarrow \pi^+\pi^-$ decays based on a $78~{\rm fb}^{-1}$
  data sample collected at the $\Upsilon(4S)$ resonance with the Belle
  detector at the KEKB asymmetric-energy $e^+e^-$ collider.  We
  reconstruct one neutral $B$ meson as a  $B^0 \rightarrow \pi^+\pi^-$
  $CP$ eigenstate and identify the flavor of the accompanying $B$
  meson from its decay products. From the asymmetry in the
  distribution of the time intervals between the two $B$ meson decay
  points, we obtain the $CP$-violating asymmetry amplitudes $\Apipi =
  \aresult$ and $\Spipi = \sresult$, where the statistical
  uncertainties are determined from Monte Carlo pseudo-experiments.
  We rule out the $CP$-conserving case, $\Apipi=\Spipi=0$, at  the
  $\cldd$ confidence level.  We discuss how these results constrain
  the  value of the CKM angle $\phi_2$.  }

\section{INTRODUCTION}
\label{sec:intro}

Kobayashi and Maskawa (KM) proposed, in 1973, a model where $CP$
violation is incorporated as an irreducible complex phase in the
weak-interaction quark mixing matrix~\cite{KM}. The KM model predicts
$CP$-violating asymmetries in the time-dependent rates for $B^0$ and
$\bzb$ decays to a common $CP$ eigenstate, $f_{CP}$, as 
\begin{eqnarray}
  \label{eq:asymmetry}
  \Acp(t)&\equiv&\!\!\frac{\Gamma(\Bzbar\to{}f_{CP};t)-\Gamma(\Bz\to{}f_{CP};t)}{\Gamma(\Bzbar\to{}f_{CP};t)+\Gamma(\Bz\to{}f_{CP};t)}
  = A_f\cos(\dmd{}t)+S_f\sin(\dmd{}t),
\end{eqnarray}
where $\Gamma(\Bz\to{}f_{CP};t)$($\Gamma(\Bzbar\to{}f_{CP};t)$) is the
partial decay rate of pure $\Bz$($\Bzbar$) state at $t=0$, and $\dmd$
is the mass difference between the two $B^0$ mass
eigenstates~\cite{Sanda}. The $CP$-violating parameters $A_f$ and
$S_f$ defined in Eq.~(\ref{eq:asymmetry}) are expressed by
$A_f=(|\lambda_f|^2 - 1)/(|\lambda_f|^2 + 1)$ and  $S_f=2Im
\lambda_f/(|\lambda_f|^2 + 1)$, where $\lambda_f$ is a complex
parameter that depends on $\bz\bzb$ mixing, on $CP$ eigenvalue of
$f_{CP}$ and on the amplitudes for $\bz$ and $\bzb$ decay to
$f_{CP}$. In the mode $\Bz\to{}J/\psi\KS$~\cite{CC}, the Standard
Model predicts $S_{J/\psi\KS}=\sin 2\phi_1$~\cite{alpha} and
$A_{J/\psi\KS}=0$. Recent measurements of the $CP$-violating parameter
$\sin 2\phi_1$ by the Belle~\cite{CP1_Belle} and
BaBar~\cite{CP1_BaBar} collaborations established $CP$ violation in
the $b\to{}c\bar{c}s$ that is consistent with KM expectations. In the
mode $\Btopipi$, we would have $\Spipi = \sin 2\phi_2$ and $\Apipi
=0$, or equivalently $|\lambda_f| = 1$, if the $b \to u$ tree
amplitude were dominant.  The situation is complicated by the
possibility  of significant contributions from gluonic $b\to d$
penguin amplitudes that have a different weak phase and additional
strong phases~\cite{pipipenguin}.  As a result,  $\Spipi$ may not be
equal to $\sin2\phi_2$ and direct $CP$ violation, $\Apipi \neq 0$, may
occur.

\section{EXPERIMENTAL APPARATUS}
\label{sec:KEKBBELLE}

In this paper, we report an updated measurement~\cite{Abe:2003ja} that
is based on a $78~{\rm fb}^{-1}$ data sample, containing
85$\times$10$^6$ $B\overline{B}$ pairs, which were collected with the
Belle detector~\cite{Belle} at the KEKB asymmetric-energy $e^+e^-$
collider~\cite{KEKB} at $\UfS$ resonance. KEKB archives the highest
peak luminosity of $8.26\times10^{33}\unilum$  as of March, 2003. The
Belle detector is a large-solid-angle general purpose spectrometer
that consists of a silicon vertex detector (SVD), a central drift
chamber (CDC), an array of aerogel threshold \v{C}erenkov counters
(ACC),  time-of-flight scintillation counters, and an electromagnetic
calorimeter comprised of CsI(Tl) crystals  located inside  a
superconducting solenoid coil that provides a 1.5~T magnetic field.
An iron flux return located outside of the coil is instrumented to
detect $K_L^0$ mesons and muons.

\section{EXPERIMENTAL METHOD}
\label{sec:expmethd}

Experimental method in this report is very similar to the
well-established $\sin2\phi_1$ measurement~\cite{CP1_Belle} in
$\Bz\to{}c\bar{c}\KS$.  In the decay chain $\Upsilon(4S)\to \bz\bzb
\to f_{CP}f_{\rm tag}$, where one of $B$ mesons decays at time
$t_{CP}$ to $f_{CP}$  and the other decays at time $t_{\rm tag}$ to a
final state $f_{\rm tag}$ that distinguishes between $B^0$ and $\bzb$,
the decay rate has a time dependence given by~\cite{CPVrev}
\begin{eqnarray}
\label{eq:R_q}
{\cal P}_{\pi\pi}^q(\Delta{t}) = \frac{e^{-|\Delta{t}|/{\taub}}}{4{\taub}} \left[1 + q\cdot \left\{ \Spipi\sin(\dmd\Delta{t}) + \Apipi\cos(\dmd\Delta{t})\right\}\right],
\end{eqnarray}
where $\taub$ is the $B^0$ lifetime, %
$\Dt$ = $t_{CP}$ $-$ $t_{\rm tag}$, and the $b$-flavor charge $q$ = +1
($-1$) when the tagging $B$ meson is a $B^0$($\bzb$). As described
below in Section~\ref{sec:pipirecon}, we reconstruct one of $B$ mesons
with two tracks identified as pion. We determine the flavor of the
accompanying $B$ from the information of its decay products. $\Dt$ is
determined from $\Dz$, the displacement in $z$ between the decay
vertices of two $B$ mesons: $\Dt \simeq (z_{CP} - z_{\rm
  tag})/\beta\gamma c \equiv\Dz/\beta\gamma c$, where $\beta\gamma$ is
a Lorentz boost factor of $0.425$. In flavor tagging and $\Dt$
measurement, we apply the same method used for the Belle $\sin
2\phi_1$ measurement~\cite{CP1_Belle3}.

The $CP$ asymmetry parameters, $\Apipi$ and $\Spipi$, are obtained
from an unbinned maximum likelihood fit to the observed $\Dt$
distribution. For this purpose, we use probability density functions,
$P_i$ (PDFs) that are based on theoretical distributions given by
Eq.(\ref{eq:R_q}). PDFs are diluted with the background and smeared by
the detector response. We use the same detector response  as those
used for the $\sin2\phi_1$
measurement~\cite{CP1_Belle3,Tajima:2003bu}.  In the fit, $\Spipi$ and
$\Apipi$ are free parameters  determined by maximizing the likelihood
function ${\cal L}=\prod_i P_i$, where the product is over all
$\Btopipi$ candidates.

\section{{$\Btopipi$} RECONSTRUCTION}
\label{sec:pipirecon}

The $\Btopipi$ event selection is described in detail
elsewhere~\cite{pipi}.   We use oppositely charged track pairs that
are positively identified as pions according to the combined
information from the ACC and the CDC $dE/dx$ measurement. The
efficiency of the pion identification is 91\% and the contamination
from miss-identified kaon is 10\% for the track momentum from 1.5 to
4.5~\unitmom. Candidate $B$ mesons are reconstructed using the energy
difference $\Delta E\equiv E_B^{\rm cms} - \Ebeam^{\rm cms}$ and the
beam-energy constrained mass $\Mbc\equiv\sqrt{(\Ebeam^{\rm
    cms})^2-(p_B^{\rm cms})^2}$, where $\Ebeam^{\rm cms}$ is the cms
beam energy, and $E_B^{\rm cms}$ and $p_B^{\rm cms}$ are the cms
energy and momentum of the $B$ candidate. The major sources of the
background in $\Btopipi$ are $\BtoKpi$ decays where kaons are
misidentified as pions, and the background from the $e^+e^-\to
q\overline{q}$ continuum ($q = u,~d,~s,~c$).

In order to suppress the continuum background, we form signal and
background likelihood functions, ${\cal L}_S$ and ${\cal L}_{BG}$,
from two variables. One is a Fisher discriminant determined from six
modified Fox-Wolfram moments~\cite{SFW}; the other is the cms $B$
flight direction with respect to the $z$ axis.  We determine ${\cal
  L}_S$ from a GEANT-based Monte Carlo (MC)
simulation~\cite{bib:Geant},  and ${\cal L}_{BG}$ from $\dE$-$\Mbc$
sideband data dominated by the continuum background. We reduce the
continuum background by imposing requirements on the likelihood ratio
$LR$ =  ${\cal L}_S/({\cal L}_S+{\cal L}_{BG})$ for candidate
events. We optimize $LR$ requirement for each flavor tagging category
to maximize the expected sensitivity.

\section{SIGNAL YIELD}
\label{sec:yield}
\begin{figure}[!htb]
\begin{center}
\resizebox{0.4\textwidth}{!}{\includegraphics{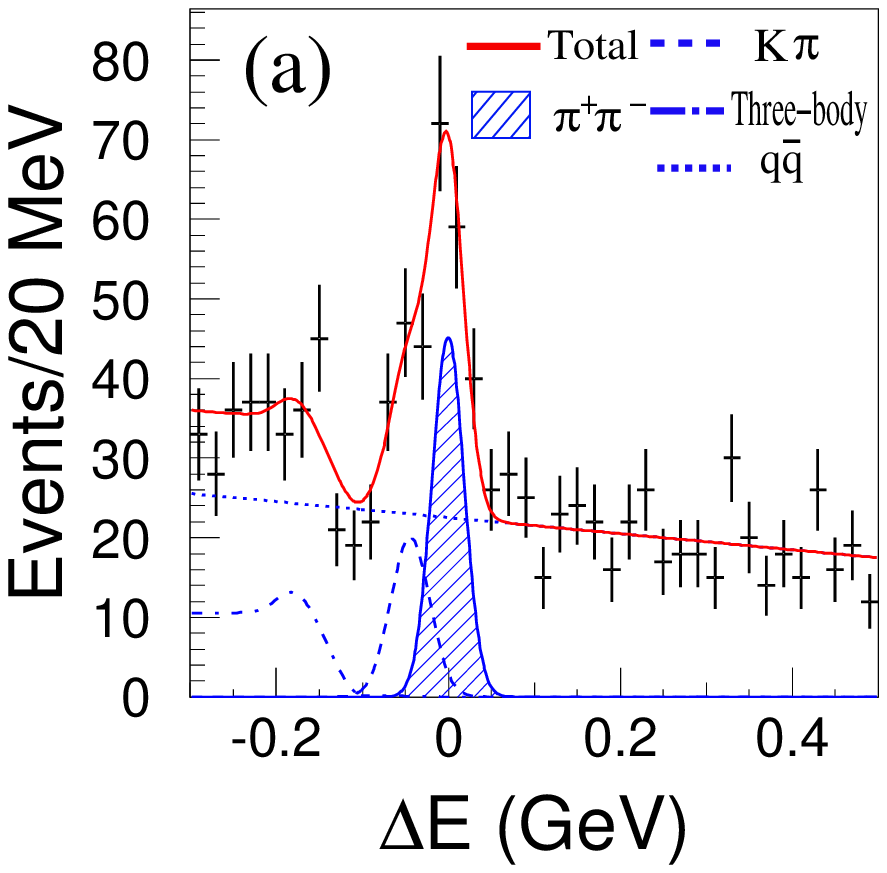}}
\resizebox{0.4\textwidth}{!}{\includegraphics{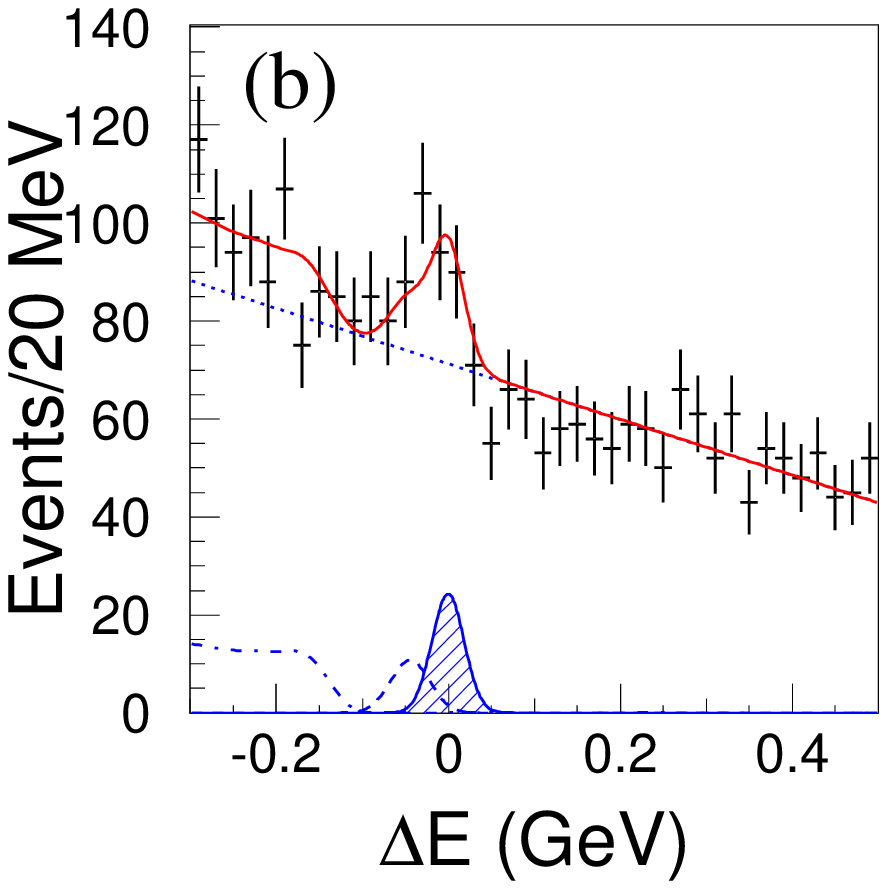}}
\end{center}
\caption{$\Delta E$ distributions in the $M_{\rm bc}$ signal region for 
(a) $B^0 \rightarrow \pi^+\pi^-$ candidates with $LR$ $>$ 0.825 and 
(b) $B^0 \rightarrow \pi^+\pi^-$ candidates with $LR$ $\leq$ 0.825.
The sum of the signal and background functions is shown as a solid curve.
The solid curve with hatched area represents the $\pi^+\pi^-$ component,
the dashed curve represents the $K^+\pi^-$ component,
the dotted curve represents the continuum background, and
the dot-dashed curve represents the charmless three-body $B$ decay
background component.}
\label{fig:DeltaE}
\end{figure}

Figures~\ref{fig:DeltaE}(a) and (b) show the $\dE$ distributions for
the $\Btopipi$ candidates that are in the $\Mbc$ signal region with
$LR$ $>$ 0.825 and with $LR$ $\leq$ 0.825, respectively,  after flavor
tagging and vertex reconstruction.  In the $\Mbc$ and $\dE$ signal
region, we find 275 candidates for $LR$ $>$ 0.825 and 485 candidates
for $LR$ $\leq$ 0.825.  The $\Btopipi$ signal yield for $LR$ $>$ 0.825
is extracted by fitting the $\dE$ distribution with a Gaussian signal
function plus contributions from misidentified $\BtoKpi$ events,
three-body $B$-decays, and continuum background. The fit yields
$106^{+16}_{-15}$ $\pi^+\pi^-$ events, $41^{+10}_{-9}$ $K^+\pi^-$
events and  $128^{+5}_{-6}$ continuum events in the signal region.
The errors do not include systematic uncertainties unless otherwise
stated. Here the error on the yield of continuum events in the signal
region is obtained by scaling the error of the yield from the fit that
encompasses the entire $\dE$ range.  For $LR\leq 0.825$,  we fix the
level of $\pi^+\pi^-$ signal by scaling the $LR>0.825$ number by a
MC-determined factor and  that of the continuum background from the
sideband.  The ratio of the $K^+\pi^-$ background to the $\pi^+\pi^-$
signal is fixed to  the value measured with the $LR> 0.825$ sample.
We obtain $57\pm 8$ $\pi^+\pi^-$ events,  $22^{+6}_{-5}$ $K^+\pi^-$
events and  $406\pm 17$ continuum events in the signal region for $LR
\leq 0.825$.  The contribution from three-body $B$-decays is
negligibly small in the signal region.

\section{FIT RESULTS}
\label{sec:result}
The unbinned maximum likelihood fit to the 760 $B^0$ $\rightarrow$
$\pi^+\pi^-$ candidates  (391 $B^0$- and 369 $\bzb$-tags),  containing
163$^{+24}_{-23}$ $\pi^+\pi^-$ signal events, yields 
\begin{eqnarray}
\Apipi &=& \aresult, \\
\Spipi &=& \sresult.
\end{eqnarray}
Here we quote the rms values of the MC $\Apipi$ and $\Spipi$
distributions as the statistical errors of our measurement, because 
the log-likelihood ratio curves from our data deviates from parabolic
behavior and the statistical error estimation using log-likelihood
ratio, $-$2ln(${\cal L}/{\cal L}_{\rm max})$, is not appropriate.

\section{CROSS CHECKS}
\label{sec:crosscheck}
We perform a number of cross checks. We measure the $B$ meson lifetime
using the same vertex reconstruction method. In addition, we check for
biases in the analysis using samples of non-$CP$ eigenstates,
$\BtoKpi$ decays, and sideband data.

We perform a $B^0$ lifetime measurement with the $\Btopipi$ candidate
events that  uses the same background fractions, vertex reconstruction
methods, and resolution functions that are used for the $CP$ fit. The
result,  $\tau_{B^0} = 1.42^{+0.14}_{-0.12}{\rm ~ps}$, is consistent
with the world-average value~\cite{PDG}. We also perform measurement of
$B^0$ lifetime and $\Bz$-$\Bzbar$ mixing with the 1371 $\BtoKpi$
candidate events (610 signal events), that are selected using
positively identified the charged kaons. we use the same vertex
reconstruction method and wrong-tag fractions as $\Btopipi$ and
determine $\taub = 1.46\pm 0.08$~ps and $\dmd =
0.55^{+0.05}_{-0.07}$~ps$^{-1}$, which are in agreement with the world
average values~\cite{PDG}.

We also use samples of non-$CP$ eigenstate $B^0 \to D^-\pi^+$,
$D^{*-}\pi^+$ and $D^{*-}\rho^+$ decays, selected with the same
event-shape criteria, to check for biases in the analysis.  The
combined fit to this control sample of 15321 events yields ${\cal A} =
-0.015 \pm 0.022$  and ${\cal S} = 0.045 \pm 0.033$. As expected,
neither mixing-induced nor direct $CP$-violating asymmetry is
observed.  A fit to the $B^0 \to K^+\pi^-$ candidates candidates
yields $\akpif = -0.03 \pm 0.11$, in agreement with the counting
analysis mentioned above~\cite{dcpv_ichep02}, and $\skpif = 0.08 \pm
0.16$, which is consistent with zero. A comparison of the event yields
and $\Delta t$ distributions for $B^0$- and $\bzb$-tagged events in
the sideband region reveals no significant asymmetry.

\section{SIGNIFICANCE}
\label{sec:CL}
We use the Feldman-Cousins frequentist approach~\cite{FeldmanCousins}
to determine the statistical significance of our measurement.  In
order to form confidence intervals, we use the $\Apipi$ and $\Spipi$
distributions of the results of fits to MC pseudo-experiments for
various input values of $\Apipi$ and $\Spipi$.  The distributions
incorporate possible biases at the boundary of the physical region as
well as a correlation between $\Apipi$ and $\Spipi$; these effects are
taken into account by this method.  The distributions are also smeared
with Gaussian functions that account for systematic
errors. Figure~\ref{fig:2dcl} shows
the resulting two-dimensional confidence regions  in the $\Apipi$
vs. $\Spipi$ plane.  The case that $CP$ symmetry is conserved,
$\Apipi=\Spipi=0$, is ruled out at the $\cldd$ confidence level
(C.L.), equivalent to $3.4\sigma$ significance for Gaussian errors.
The minimum confidence level for $\Apipi$ = 0, the case of no direct
$CP$ violation, occurs  at $(\Spipi, \Apipi$) = ($-1.0, 0.0)$ and is
97.3$\%$, which corresponds to 2.2$\sigma$ significance.

\section{DISCUSSION}
\label{sec:phi2}

\begin{figure}[!htb]
  \begin{center}
    \resizebox{0.5\textwidth}{!}{\includegraphics{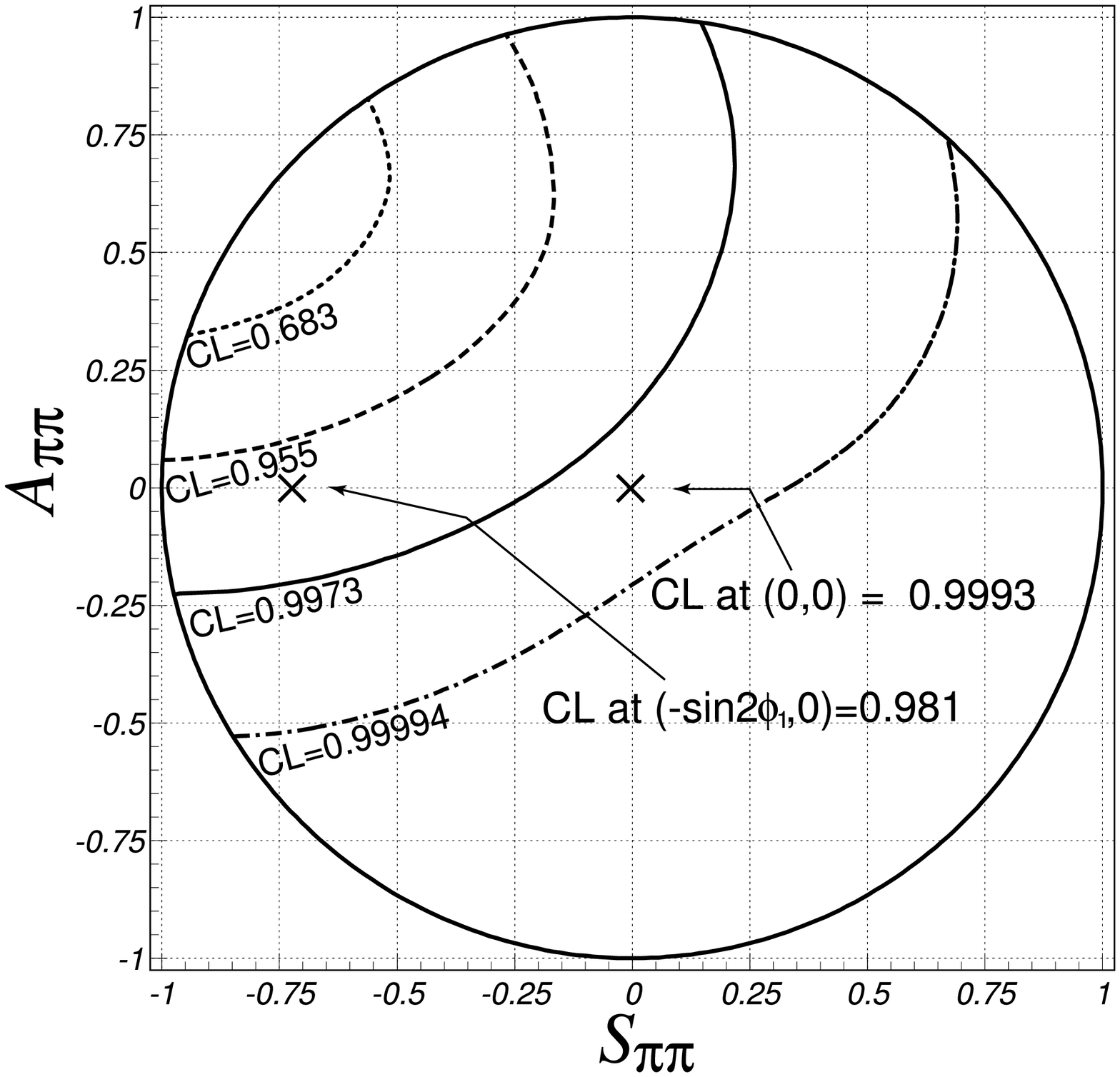}}
  \end{center}
\caption{Confidence regions for $\apipi$ and $\spipi$.}
\label{fig:2dcl}
\end{figure}

\begin{figure*}[!htbp]
\resizebox{0.85\textwidth}{!}{\includegraphics{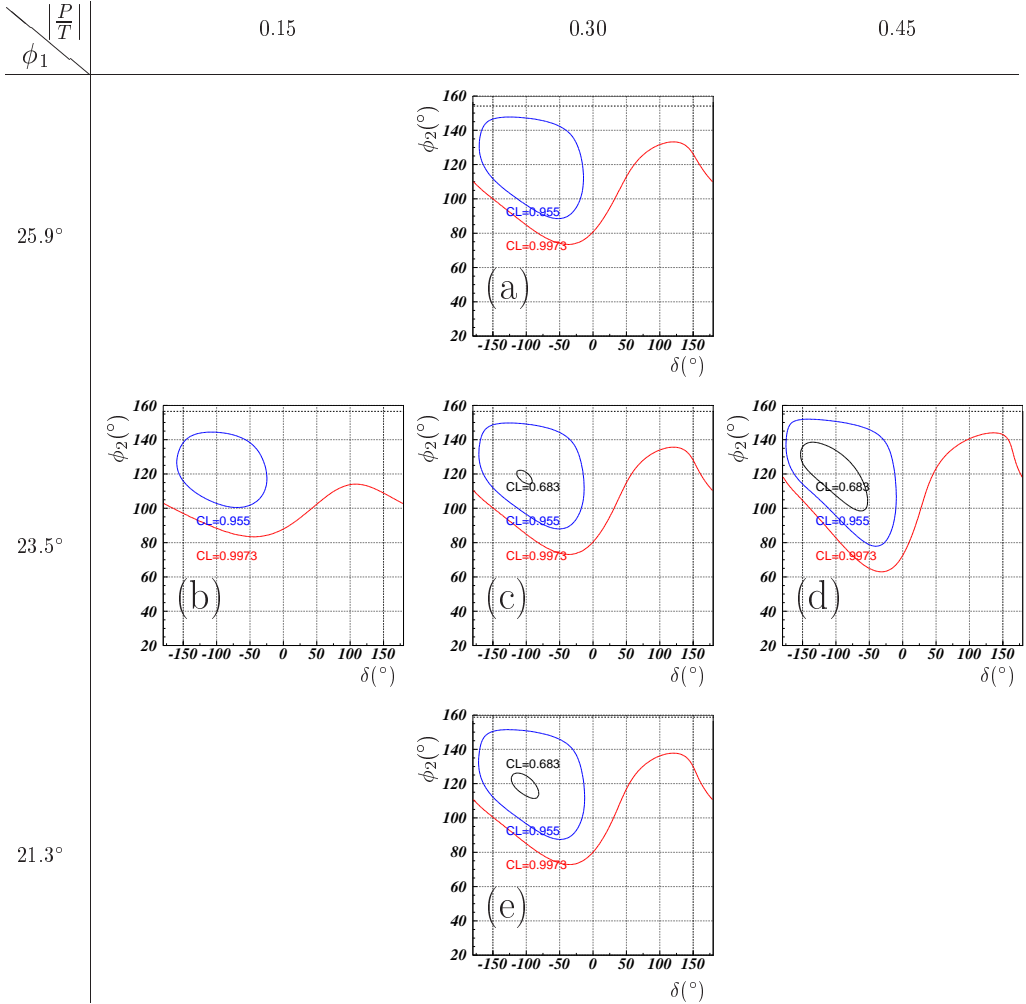}}
\caption{
The regions for $\phi_2$ and $\delta$ corresponding to the 68.3$\%$, 95.5$\%$, and 
99.73$\%$ C.L. regions of $\apipi$ and $\spipi$ in Fig.~\ref{fig:2dcl} for
(a) $\phi_1=25.9^\circ$, $|P/T|$=0.3, (b) $\phi_1=23.5^\circ$, $|P/T|$=0.15,
(c) $\phi_1=23.5^\circ$, $|P/T|$=0.3, (d) $\phi_1=23.5^\circ$, $|P/T|$=0.45,
and (e) $\phi_1=21.3^\circ$, $|P/T|$=0.3. The horizontal dashed lines correspond to $\phi_2 = 180^\circ - \phi_1$.}
\label{fig:phi2-delta}
\end{figure*}

Using the standard definitions of weak phases $\phi_1$, $\phi_2$, and $\phi_3$,
the decay amplitudes for $B^0$ and $\overline{B}^0$ to $\pi^+\pi^-$ are
\begin{eqnarray}
A(B^0\rightarrow\pi^+\pi^-) &=& -(|T|e^{i\delta_{T}}e^{i\phi_3}~~ +
|P|e^{i\delta_P}), \\
A(\overline{B}^0\rightarrow\pi^+\pi^-) &=& -(|T|e^{i\delta_{T}}e^{-i\phi_3} +
|P|e^{i\delta_P}) ,
\end{eqnarray}
where $T$ and $P$ are the amplitudes for the tree and penguin 
graphs and $\delta_T$ and $\delta_P$ are their strong phases. 
Here we  adopt the notation of Ref.~\cite{bib:SA_TH_GR} and use
the convention in which the top-quark contributions are integrated out
in the short-distance effective Hamiltonian. In addition, the unitarity relation
$V^*_{ub}V_{ud}$ + $V^*_{cb}V_{cd}$ = $-V^*_{tb}V_{td}$ is applied. 
Using the above expressions and $\phi_2$ = $\pi - \phi_1 - \phi_3$, we determine
$\lambda_{\pi\pi}\equiv{}e^{2i\phi_2}[1+|P/T|e^{i(\delta+\phi_3)}]/[1+|P/T|e^{i(\delta-\phi_3)}]$.
Explicit expressions for $\Spipi$ and $\Apipi$ are
\begin{eqnarray}
\Spipi &=& [{\rm sin}2\phi_2 + 2|P/T|{\rm sin}(\phi_1 - \phi_2){\rm cos}{\delta}- |P/T|^2{\rm sin}2\phi_1]/\rpipi, \\
\Apipi &=& - [2|P/T|{\rm sin}(\phi_2 + \phi_1){\rm sin}{\delta}]/\rpipi, \\
\rpipi &=& 1 - 2|P/T|{\rm cos}{\delta}{\rm cos}(\phi_2 + \phi_1) + |P/T|^2,
\end{eqnarray}
where $\delta$ $\equiv$ $\delta_P$ $-$ $\delta_T$.  We take
$-180^\circ\leq\delta\leq{180}^\circ$.  When $\Apipi$ is positive and
$0^\circ<\phi_1+\phi_2<180^\circ$, $\delta$ is  negative. Recent
theoretical estimates  prefer $|P/T| \sim 0.3$ with large
uncertainties~\cite{bib:GR,bib:LR,bib:Beneke,bib:phi2-th}.
Figures~\ref{fig:phi2-delta}(a)-(e) show the regions for $\phi_2$ and
$\delta$   corresponding to the 68.3$\%$ C.L., 95.5$\%$ C.L. and
99.73$\%$ C.L. region of $\Apipi$ and $\Spipi$   (shown in
Fig.~\ref{fig:2dcl}) for representative values of $|P/T|$ and
$\phi_1$~\cite{bib:phi1}.   Note that a value of ($\Spipi$,$\Apipi$)
inside the 68.3$\%$ C.L. contour requires a value of $|P/T|$ greater
than $\sim$0.3. The allowed region is not very sensitive to variations
of $\phi_1$ within the errors of the measurements, as can be seen by
comparing Figs.~\ref{fig:phi2-delta}(a), (c) and (e).  The range of
$\phi_2$ that corresponds to the 95.5$\%$ C.L. region of  $\Apipi$ and
$\Spipi$ in Fig.~\ref{fig:2dcl} is
\begin{eqnarray}
78^\circ \leq \phi_2 \leq 152^\circ,
\end{eqnarray}
for $\phi_1=23.5^\circ$ and $0.15\leq|P/T|\leq0.45$.
The result is in agreement with constraints on the unitarity triangle
from other measurements~\cite{bib:NIR02}.

\section{CONCLUSION}
\label{sec:conclusion}
In summary, we have performed an improved measurement of $CP$
violation parameters in $B^0 \rightarrow \pi^+\pi^-$ decays.  An
unbinned maximum likelihood fit to  760 $B^0$ $\rightarrow$
$\pi^+\pi^-$ candidates, which contain $163^{+24}_{-23}$(stat)
$\pi^+\pi^-$ signal events, yields $\Apipi = \aresult$, and $\Spipi =
\sresult$, where the statistical uncertainties are determined from MC
pseudo-experiments.  This result is consistent with our previous
measurement~\cite{Acp_pipi_Belle} and supersedes it.  We obtain
confidence intervals for $CP$-violating asymmetry parameters $\Apipi$
and $\Spipi$ based on the Feldman-Cousins approach  where we use MC
pseudo-experiments to determine acceptance regions.  We rule out the
$CP$-conserving case, $\Apipi=\Spipi=0$, at  the $\cldd$ confidence
level.

The result for $\Spipi$ indicates that mixing-induced $CP$ violation
is large, and the large $\Apipi$ term is an indication of direct $CP$
violation in $B$ meson decay.  Constraints within the Standard Model
on the CKM angle $\phi_2$ and the hadronic phase difference between
the tree ($T$) and penguin ($P$) amplitudes are obtained for $|P/T|$
values that are favored theoretically.  We find an allowed region of
$\phi_2$ that is consistent with constraints on the unitarity triangle
from other measurements.

\begin{figure}[!htbp]
\begin{center}
\resizebox{0.6\textwidth}{!}{\includegraphics{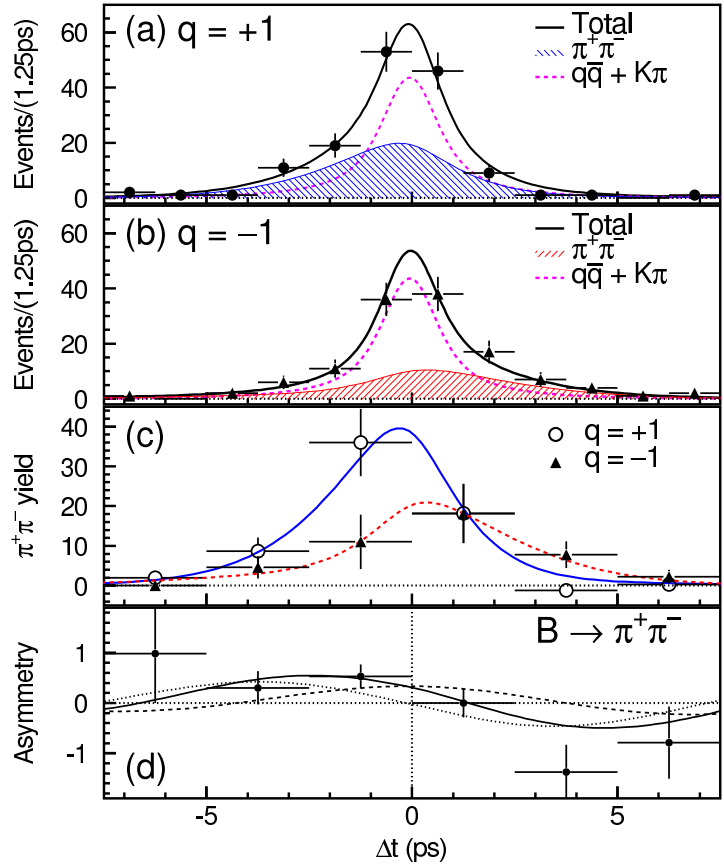}}
\end{center}
\caption{
The raw, unweighted $\Delta t$ distributions for 
the 275 $B^0 \rightarrow \pi^+\pi^-$ candidates with $LR$ $>$ 0.825 in the
signal region: 
(a) 148 candidates with $q = +1$, i.e. the tag side is identified as $B^0$;
(b) 127 candidates with $q = -1$; 
(c) $B^0$ $\rightarrow$ $\pi^+ \pi^-$ yields after background subtraction. The errors
are statistical only and do not include the error on
the background subtraction;
(d) the $CP$ asymmetry for $B^0 \rightarrow \pi^+\pi^-$
after background subtraction.
In Figs. (a) through (c), the curves show the
results of the unbinned maximum likelihood fit to the $\Delta t$
distributions of the 760 $B^0$ $\rightarrow$ $\pi^+\pi^-$ candidates.
In Fig. (d), the solid curve shows the resultant $CP$ asymmetry,
while the dashed (dotted) curve is the contribution from
the cosine (sine) term.
}
\label{fig:asym}
\end{figure}

\begin{thebibliography}{99}
\bibitem{CC} 
  Throughout this paper,  the inclusion of the charge conjugate mode
  decay is implied unless otherwise stated.
\bibitem{bib:phi1}
Using the average of recent values of sin2$\phi_1$ 
from Belle~\cite{CP1_Belle3} and BaBar~\cite{CP1_BaBar},
we obtain $\phi_1=(23.5^{+2.4}_{-2.2})^\circ$.
\bibitem{Acp_pipi_Belle}
Belle Collaboration, K.~Abe \textit{et al.}, \Journal{\PRL}{89}{071801}{2002}.
$\spipif=-1.21^{+0.38}_{-0.27}$(stat)$^{+0.16}_{-0.13}$(syst) and 
$\apipif=+0.94^{+0.25}_{-0.31}$(stat)$\pm{0.09}$(syst).
In this paper, the statistical uncertainties were determined from the observed
 fitting errors rather than from MC pseudo-experiments as in the analysis
 described here. 
Thus, the errors here are larger than those obtained by scaling the previous errors by the integrated luminosities.
\end{thebibliography}
\end{document}